\documentstyle[twocolumn,eqsecnum,aps,epsfig]{revtex}
\def\eqcomma{\quad ,}
\def\half{\mbox{$\frac{1}{2}$}}     
\def\eqperiod{\quad .}
\begin{document}
\preprint{UMD PP-00-029; gr-qc/9910032}
\draft
\twocolumn[\hsize\textwidth\columnwidth\hsize\csname
@twocolumnfalse\endcsname
\title{Relativistic Scalar Gravity:\\
A Laboratory for Numerical Relativity%
\protect\cite{PP}
}
\author{Keith Watt and Charles W. Misner}
\address{
University of Maryland\\
College Park, MD 20742\\
email: {\tt kwatt@astro.umd.edu, misner@umail.umd.edu}
}
\date{9 October 1999}
\maketitle
\widetext
\begin{abstract}
We present here a relativistic theory of gravity in which the 
spacetime metric is derived from a single scalar field $\Phi$.  
The field equation, derived from a simple variational principle, is a 
non-linear flat-space four-dimensional wave equation which is 
particularly suited for numerical evolution. 
We demonstrate that while this theory does not generate results 
which are exactly identical quantitatively to those of general 
relativity (GR), many of the qualitative features of the full GR theory 
are reproduced to a reasonable approximation.  
The advantage of this formulation lies in the fact that 3D numerical 
grids can be numerically evolved in minutes or hours instead of the 
days and weeks required by GR, thus drastically reducing the 
development time of new relativistic hydrodynamical codes.  
Scalar gravity therefore serves as a meaningful testbed for the 
development of larger routines destined for use under the full theory 
of general relativity.
\end{abstract}

\narrowtext
\vskip2pc]

\section{Scalar Theories of Gravity}
\label{sec:ScalarTheories}

Computational loads in numerical relativity are typically immense.  
Fox \cite{Fox96} estimates that the computation of the gravitational 
waves from black hole collisions would require approximately 
100,000 hours on a Cray Y-MP.  
Current \cite{SeidSu99} improvements in parallel
machines and algorithms may cut this to 50 hours on rare
machines.
It is very difficult to do even test problems under these conditions. 
Clearly, one cannot wait days or weeks to get results to use in 
deciphering the cause (or even detecting the existence) of some 
problem in the code. 
When one considers that we are not even completely certain what 
new problems we will be forced to face, the situation is completely 
unacceptable. 
While many problems will be unique to the Einstein equations and
must be faced in that context, others, notably in relativistic
hydrodynamics, will be similar using any metric theory of
gravity.

Shapiro and Teukolsky \cite{ST93} have proposed that instead of 
starting with the full equations of general relativity, we should 
instead look at a simpler relativistic theory of gravity in order to 
discover and learn how to resolve these unknown issues while 
utilizing far less computational resources than the full problem 
would require. 
The scalar theory presented here is intended to provide a testbed
for relativistic hydrodynamics.
It is to be implemented in code that accepts a matter 
stress-energy tensor and generates an updated metric at each 
time step.
The hydrodynamic code that accepts this metric and updates the
hydrodynamic variables and the stress-energy tensor can be
exactly the same code that is to interact in the same way with a
program that solves the Einstein equations.
Thus speeded development and testing of hydrodynamical codes is
the principal motivation for developing this scalar theory.
Any gravitational physics or techniques that may be carried over
to Einstein equation solvers will be a bonus.

Shapiro and Teukolsky's prescription is to explore the scalar 
gravitation theory presented in Exercise 7.1 of Misner, Thorne, and 
Wheeler (MTW) \cite{MTW73}. 
This theory, however, violently disagrees with observations and fails 
all three of the classical tests of general relativity. 
It is not clear that this theory is close enough to general relativity to 
prove useful to our purposes. Nevertheless, the potential benefits of 
a simplified testbed are such that a further examination of scalar 
gravity theories is warranted.

Ni \cite{Ni72} provides a compendium of metric theories organized 
in terms of their ``Parameterized Post-Newtonian limits" as defined 
by Thorne and Will \cite{ThorneWill71}. 
In general, scalar theories can be grouped into three classifications:
\begin{enumerate}
	\item Conformally flat theories
	\item Stratified conformally flat (SCF) theories
	\item Other scalar theories
\end{enumerate}
Conformally flat theories are some sense the simplest possible 
theories of gravity.  
In this class of theories, the metric takes the form
\begin{equation}
\label{eq:CFGenMetric}
ds^{2}=\Psi(-dt^{2}+dx^{2}+dy^{2}+dz^{2}) \eqcomma
\end{equation}
where the variable $\Psi$ is some function of the scalar field $\Phi$, 
e.g., $\Psi = \exp{2\Phi}$.  
We use the symbol $\Phi$ throughout for the basic scalar 
gravitational field, normalized so that in the Newtonian limit it 
becomes the Newtonian gravitational potential, asymptotically 
$\Phi = -M/r = -GM/rc^2$.
We shall also use units with $G = 1 = c$ where $G$ is the 
Newton-Cavendish gravitational constant.
The field equation (which determines $\Phi$) can technically be 
anything, but is generally some variation of the scalar wave 
equation, with some version of the matter density serving as the 
source term.  
Ni describes five different examples of this type of theory. 
\cite{Ni72} 

All conformally flat theories (including the theory of MTW Ex.7.1) 
suffer from a fatal flaw: since Maxwell's equations are conformally 
invariant, all conformally flat theories must predict zero light 
deflection, in obvious disagreement with experiment.  
Most also fail various other solar-system tests of general relativity, 
but the zero light-deflection prediction is a fundamental 
consequence of each of these theories and cannot be avoided.  

One way to solve the light-deflection problem is to postulate that 
there exists a universal preferred reference frame and that while 
spacelike slices of this spacetime (referred to as ``strata") are 
conformally flat, the full spacetime is not.  
The general form for the metric of this ``stratified conformally flat 
theory" is
\begin{equation}
\label{eq:SCFGenMetric}
ds^{2}=-\Psi_{1}dt^{2}+\Psi_{2}(dx^{2}+dy^{2}+dz^{2}) \eqcomma
\end{equation}
where again $\Psi_{1}$ and $\Psi_{2}$ are functions of the scalar 
field $\Phi$.  
This type of metric obviously destroys Lorentz invariance, however 
this is not as strict a requirement as one might think.  
Global Lorentz invariance is a treasured philosophical feature of 
general relativity with asymptotically flat boundary conditions, but in 
reality we usually model sources as though they existed alone and 
in isolation in the Universe. 
The fact that we have a preferred reference frame in this case is not 
such an important issue.   

The last category, more general scalar theories, has seen little 
examination in the literature and we do not study it here.
 
\section{Simple Relativistic Scalar Gravity}
\label{sec:SimpleRSG}

\subsection{Variational Principle and Metric}
\label{sec:VarPrin}

We propose a variational principle which leads to a simplified theory 
closely related to Ni's \cite{Ni72} ``Lagrangian-based stratified 
theory", which until 1972 was thought to agree with all tests of 
general relativity. 
It has since been found to disagree with the Earth rotation rate 
experiment \cite{NordtvedtWill72} and with white dwarf stability
observations 
\cite{Ni73} (although Ni notes that the validity of the latter 
experiment can be questioned since it is based upon ideas about 
pulsation-damping forces in white dwarfs which have yet to be 
tested experimentally). 
As will be seen in Section \ref{subsec:PPN}, the PPN form of the 
simplified metric is identical to that found from Ni's theory, hence it 
satisfies (and fails) the same tests of general relativity as does Ni's 
theory.  
This stratified conformally-flat (SCF) scalar theory, therefore, has all 
of the advantages of Ni's more complicated theory, yet has a 
simpler form which eliminates many of the disadvantages.  
Additionally, the resulting field equation has the form of the 
non-linear Klein-Gordon equation, for which a significant amount 
literature exists, thus aiding in the numerical implementation of the 
theory.  

We begin by selecting the Lagrangian:  
\begin{equation}
\label{eq:SCFVarPrin}
I=-\frac{1}{8 \pi} \int \eta^{\mu \nu} \Phi_{, \mu} \Phi_{, \nu} 
\sqrt{-\eta}\, d^{4}x + \int 
{\cal L}_{\rm matter} d^{4}x \eqperiod
\end{equation} 
This differs from Ni's theory (cf. \cite{Ni72} or a summary in MTW 
\cite[Box 39.1]{MTW73}) only in using $\eta^{\mu \nu}$ where Ni 
uses $g^{\mu \nu}$ in the first term, so that our scalar field $\Phi$ 
propagates along flat spacetime light cones.  
Here $\eta^{\mu\nu}$ is the Minkowski metric, and $\sqrt{-\eta} = 
1$ unless one is using nonstandard, {\it e.g.}, spherical, coordinates.
We generate the metric (with which all nongravitational fields interact 
just as in general relativity) in the same way as Ni:
\begin{equation}
\label{eq:SCFmetric}
ds^{2}=-e^{2 \Phi}dt^{2}+e^{-2 \Phi}(dx^{2}+dy^{2}+dz^{2}) 
\eqperiod
\end{equation}
We have briefly explored alternative forms of the stratified metric,
\begin{equation}
\label{eq:SCFnformmetric}
ds^{2}=-(1-\frac{2}{n} \Phi)^{-n}dt^{2}+(1-\frac{2}{n} \Phi)^{n}
(dx^{2}+dy^{2}+dz^{2}) ,
\end{equation}
but found that the most realistic ({\it i.e.}, closest to Einstein) 
results are obtained when $n$ is allowed to go to infinity.

\subsection{The Field Equation}
\label{subsec:FieldEquation}

The field equation follows from varying $\Phi$ in the action 
(\ref{eq:SCFVarPrin}) which gives
\begin{equation}
\label{eq:FieldEqA}
\frac{\partial}{\partial x^\mu}(\sqrt{-\eta}\, \eta^{\mu \nu}
\frac{\partial \Phi}{\partial x^\nu}) 
= - 4 \pi \frac{\delta {\cal L}_{\rm matter}}{\delta g_{\mu\nu}} 
\frac{\partial g_{\mu\nu}}{\partial \Phi} 
\eqperiod
\end{equation}
The right hand side of this equation can be expressed in terms of 
the stress-energy tensor by using the definition found in MTW 
\cite[Chap. 21]{MTW73}):
\begin{equation}
\label{eq:Tdefn}
\frac{\delta {\cal L}_{\rm matter}}{\delta g_{\mu \nu}}  
= \half T^{\mu \nu}\sqrt{-g}
\eqperiod
\end{equation}
The field equation derived from this variational principle, which is 
valid for a metric generated in any way from $\Phi$, is therefore
\begin{equation}
\label{eq:GenFieldEq}
\sqrt{-\eta}\, \Box \Phi = - 2 \pi \sqrt{-g} 
\frac{\partial g_{\mu \nu}}{\partial \Phi} T^{\mu \nu} \eqcomma
\end{equation}
where $\sqrt{-\eta}\, \Box = \partial_\mu \sqrt{-\eta}\,
\eta^{\mu \nu} \partial_\nu$.  
Utilizing the SCF metric (\ref{eq:SCFmetric}), this field 
equation becomes
\begin{equation}
\label{eq:SCFFieldEq}
\sqrt{-\eta}\, \Box \Phi 
=  4 \pi (T^{00}+e^{-4 \Phi} \delta_{ij} T^{ij}) \eqperiod
\end{equation}
The theory proposed by Ni \cite{Ni72} differs only in using the 
metric (\ref{eq:SCFmetric}) where our simplified theory uses the 
Minkowski metric $\eta_{\mu\nu}$ on the left hand side.  

\subsection{Conserved Quantities}
\label{subsec:ConservedQuantities}

It is often important to know which physical quantities are 
conserved in a theory.  
These quantities are of particular importance if the field equation is 
to be solved numerically, as they are often used as to monitor the
error or stability of a numerical integration algorithm
(e.g. \cite{NewWatt98}).
A conserved energy and conserved linear and angular momentum 
can be derived from the Lagrangian implicit in the variational 
principle of (\ref{eq:SCFVarPrin}) using Noether's Theorem, 
since this Lagrangian density is invariant under spacetime
translations and spatial rotations.
Expositions of Noether's theorems can be found, e.g., in
Hill \cite{Hill51} and in Byers \cite{Byers98}.
For the present application the special case using the canonical
stress-energy tensor suffices.  For a Lagrangian density ${\cal
L}(\Psi^{\sc A},{\Psi^{\sc A}}_{,\mu})$ which has no explicit
coordinate dependence but may contain several
fields $\Psi^{\sc A}$, the equations 
\begin{equation}
\label{eq:SGConsvEq1}
{{\cal T}_{\lambda}}^{\gamma} _{,\gamma} =  0 
 \end{equation}
where
\begin{equation}
\label{eq:SGConsvTcanonical}
 {{\cal T}_{\lambda}}^{\gamma} 
  =  {\cal L}\delta_\lambda^\gamma 
 - \frac{\partial {\cal L}}{\partial {\Psi^{\sc A}}_{,\gamma}}
    {\Psi^{\sc A}}_{,\lambda}
 \end{equation}
are easily shown to be a consequence of the field equations
\begin{equation}
\label{eq:SGConsvFieldEqs}
 0  =  \frac{\partial{\cal L}}{\partial {\Psi^{\sc A}}} 
 - \partial_\gamma
\frac{\partial {\cal L}}{\partial {\Psi^{\sc A}}_{,\gamma}} 
 \eqperiod
 \end{equation}
(See, e.g., \cite{LL71,Ryder85,PeskSch95}.)
Since the stress energy tensor ${{\cal T}_\lambda}^\gamma$ as
defined in equation~(\ref{eq:SGConsvTcanonical}) is linear in 
$\cal L =  L_{\rm grav} +  L_{\rm matter}$, it
is also a sum of two parts and the energy-momentum conservation 
law can be written as 
\begin{equation}
\label{eq:SGConsvEq2}
 \left[{{\cal T}_{({\rm grav})\lambda}}^{\gamma} 
 + {{\cal T}_{({\rm matter})\lambda}}^{\gamma}\right]
 _{,\gamma} =  0 \eqperiod
 \end{equation}
 Here
 \begin{equation}
 \label{eq:TDefnGrav}
 {{\cal T}_{({\rm grav})\lambda}}^{\gamma} = \frac{1}{4 \pi}
 \sqrt{-\eta}\left( -\half \eta^{\mu \nu} \Phi_{, \mu} \Phi_{, \nu}\,
 \delta^{\gamma}_{\lambda} +  \Phi_{, \lambda} \Phi_{, \mu} 
 \eta^{\mu \gamma} \right) 
 \end{equation}
follows from equation~(\ref{eq:SGConsvTcanonical}) since $\cal 
L_{\rm grav}$ depends on only the single field $\Phi$.
Although $\cal L_{\rm matter}$ may depend on several matter
fields, in it $\Phi$ occurs only indirectly via $g_{\mu\nu}$. 
Since we assume that $\cal L_{\rm matter}$ is generally covariant,
one knows that its stress-energy tensor can equivalently be
obtained as
\begin{equation}
 \label{eq:TDefnMatter}
 {{\cal T}_{({\rm matter})\lambda}}^{\gamma} = 2 g_{\lambda\alpha}
  \frac{\delta{\cal L}_{\rm matter}}{\delta g_{\alpha\gamma}} 
  = g_{\lambda\alpha} \sqrt{-g}\, T^{\alpha\gamma}
  \eqperiod
 \end{equation}
To directly verify the conservation law (\ref{eq:SGConsvEq2}) 
one uses the scalar field equation (\ref{eq:SCFFieldEq}) and 
the covariant conservation law for the matter fields 
$ {T^{\lambda\gamma}}_{;\gamma} =  0 $ in the form
\begin{eqnarray}
  {{{\cal T}_{({\rm matter})\lambda}}^{\gamma}}
 _{,\gamma} & = & \Gamma_{\alpha\lambda\gamma}
                      {{\cal T}_{({\rm matter})}}^{\alpha\gamma}
                      \nonumber \\
            & = & \half g_{\alpha\gamma,\lambda} \sqrt{-g}\,
                        T^{\alpha\gamma}
                       \\
            & = &  -\Phi_{,\lambda}(T^{00}+e^{-4 \Phi} \delta_{ij} T^{ij}) 
                   \eqperiod   \nonumber 
 \end{eqnarray}
Only the last step in this equation uses the special SCF form 
(\ref{eq:SCFmetric}) of $g_{\mu\nu}$ in terms of $\Phi$.

Since the $\eta^{\mu\nu}$ are constants, the conservation law
${{{\cal T}_\lambda}^\gamma}_{,\gamma}=0$ can also be written
${{\cal T}^{\lambda\gamma}}_{,\gamma}=0$ by defining
${{\cal T}^{\lambda\gamma}} = \eta^{\lambda\alpha}{{{\cal T}_\alpha}^\gamma}$.
One then defines the total energy and momentum of the system as usual by
\begin{equation} 
\label{eq:4MomentumDefn}
  P^\mu = \int {{\cal T}^{\mu 0}}\,d^3x
\end{equation}
and recognizes ${\cal T}^{0 k}$ as the Poynting-like energy flux.
The zero-component of (\ref{eq:4MomentumDefn}) is the energy, 
defined as
\begin{equation}
\label{eq:EnergyDefn}
  E \equiv P^{0} = -\int {{\cal T}_{0}^{0}}\,d^3x
 \eqcomma
\end{equation}
with the corresponding energy conservation law
\begin{equation}
\label{eq:EnergyConsv}
\frac{dE}{dt} = -\oint {{\cal T}^{0 k}}\,d^2 \sigma_k 
\eqperiod
\end{equation}
Here ${\cal T}^{0 k} = - {{\cal T}_{0}}^k \equiv {{\cal S}^{k}}$ 
is an energy flux like the Poynting vector.  In most 
applications the bounding surface will be chosen outside the 
region occupied by matter so only the gravitational terms remain 
as ${\cal S}^{k}_{\rm grav} = - (c^5/4\pi G)\Phi_{,0}\Phi_{,k}$.

Angular momentum conservation is a consequence of the symmetry 
of the stress-energy tensor.  
The gravitational part of ${\cal T}^{\lambda\gamma}$ is 
symmetric as a consequence of the Lorentz invariance of the 
Lagrangian ${\cal L}_{\rm grav}$ from which it is derived, and 
is also clear from its explicit definition
\begin{equation}
 \label{eq:TSymmGrav}
 {\cal T}_{({\rm grav})}^{\lambda\gamma} = \frac{1}{4 \pi}
 \sqrt{-\eta}\left( -\half \eta^{\mu \nu} \Phi_{, \mu} \Phi_{, \nu}\,
 \delta^{\lambda\gamma} +  \eta^{\lambda\mu} \Phi_{, \mu} \Phi_{, \nu} 
 \eta^{\nu \gamma} \right) \ \ .
\end{equation}
However the matter Lagrangian is not Lorentz invariant in view of
the preferred rest frame in which equation~(\ref{eq:SCFmetric})
holds.  The matter stress-energy is defined by
\begin{equation}
 \label{eq:TSymmMatter}
 {{\cal T}_{({\rm matter})}}^{\lambda\gamma} = 
  2 \eta^{\lambda\beta} g_{\beta\alpha}
  \frac{\delta{\cal L}_{\rm matter}}{\delta g_{\alpha\gamma}} 
  = \eta^{\lambda\beta} g_{\beta\alpha} \sqrt{-g}\, T^{\alpha\gamma}
 \end{equation}
and does not inherit the symmetry of $T^{\alpha\beta}$ except in
its spatial components:  
${{\cal T}_{({\rm matter})}}^{ij} = 
{{\cal T}_{({\rm matter})}}^{ji} = e^{-4\Phi}T^{ij}$ .
An angular momentum density can be defined as usual
${\cal M}^{\mu\nu\gamma} = x^\mu {\cal T}^{\nu\gamma}
                  - x^\nu {\cal T}^{\mu\gamma}$
which satisfies
${{\cal M}^{\mu\nu\gamma}}_{,\gamma} = {\cal T}^{\nu\mu} 
                                      - {\cal T}^{\mu\nu}$
so that a conserved angular momentum is defined
\begin{equation} 
  J^{ij} = \int {{\cal M}^{i j 0}}\,d^3x
  \eqperiod
\end{equation}

\subsection{The Newtonian Limit}
\label{subsec:NewtonianLimit}

In order to find the Newtonian limit for this theory, assume a static 
point source of mass $M$ for the stress-energy tensor and solve 
the field equation for the Schwarzschild-like solution.  
The field equation with this source is
\begin{equation}
\label{eq:SchwFieldEq}
\bigtriangledown^{2} \Phi = 4 \pi M \delta (x) \eqperiod
\end{equation}
Away from the source, the solution to this equation is
\begin{equation}
\label{eq:SchwSolution}
\Phi = -\frac{M}{r} \eqcomma
\end{equation}
which is, of course, the Newtonian potential.  The spacetime metric 
is therefore
\begin{equation}
\label{eq:SchwMetric}
ds^{2}=-e^{-{2 M}/{r}}dt^{2}+e^{{2 M}/{r}}
(dx^{2}+dy^{2}+dz^{2}) \eqperiod
\end{equation}
The small-$\Phi$ (small $M$ or large $r$) limit of the $g_{00}$ 
component is $g_{00}\rightarrow -(1+2\Phi)$, therefore 
correspondence with Newtonian gravity is assured.

\subsection{PPN Parameters}
\label{subsec:PPN}

Having found the field equation, one is now compelled to ask, how 
``good" is the theory?  
We will explore a number of definitions of ``good" in this paper, but 
one possible definition is to insist that a ``good" theory have 
parameterized post-Newtonian (PPN) parameters which 1) satisfy 
as many as possible of the solar system gravitational tests made to 
date, and 2) match as closely as possible the PPN parameters of 
general relativity.  
One must keep in mind, however, that obviously a theory may be 
``good" in its Newtonian limit yet still give results which differ greatly 
from general relativity in the strong-field limit!
In fact, we have found no solutions in this SCF scalar theory
that approximate the cosmological models in general relativity.

Ni \cite{Ni72} has calculated the PPN parameters for his
 ``Lagrangian-based stratified theory", which our simplified theory
 closely resembles.  
 Subtracting the present field equation (\ref{eq:SCFFieldEq}) from 
 Ni's shows that the theories differ by
  \begin{eqnarray}
 \label{eq:PPNComparison}
 (\sqrt{-g} \, & g^{\mu \nu}&)_{,\mu} \Phi_{,\nu}
 - (\sqrt{-\eta} \, \eta^{\mu \nu})_{,\mu} \Phi_{,\nu}
 \nonumber \\
 &=& 4 e^{-4 \Phi} (\Phi_{,0})^2 + (1-e^{-4 \Phi}) \Phi_{,00}
 \eqperiod
 \end{eqnarray}
 Both of these terms are $O(c^{-6})$. Because the PPN formulation 
 considers terms through $O(c^{-4})$, this theory and Ni's must have 
 identical PPN parameters. 
 [The PPN approximation is often considered an expansion in
powers of $1/c$ since the dimensionless quantities $(v/c)^2$ and
$GM/rc^2$ are of comparable size (virial theorem) in bound
systems, and timelike derivatives $\partial_0$ are smaller by 
a factor of order $v/c$ than space derivatives $\partial_i$ when
the time dependence is generated by the motion of sources.
The Newtonian potential $\Phi \sim -GM/rc^2$ is considered
to be $O(c^{-2})$.]
 The PPN parameters for this theory \cite{Ni72} are therefore
\begin{eqnarray}
\label{eq:PPNparams}
\xi=\eta=0, \gamma=\beta=\beta_{1}&=&\beta_{2}=\beta_{3}
=\beta_{4}=\Delta_{2}=1, \nonumber \\
\Delta_{1}&=&-\frac{1}{7} \eqperiod
\end{eqnarray}

The theory agrees with general relativity with the exception of the 
$\Delta_{1}$ parameter.  
Additionally, due to its value of $\Delta_{1}$, it disagrees 
with the Earth rotation rate experiments. \cite{NordtvedtWill72}  
Essentially, this theory predicts an incorrect relative amount of 
dragging of inertial frames ($g_{0i}$) produced by unit momentum 
($\rho v$) (see MTW Box 39.2 \cite{MTW73}).  
On the basis of PPN parameters, this theory compares favorably to 
all of its scalar gravity competitors.  
In fact, one obtains precisely the PPN parameters of Ni's SCF 
theory \cite{Ni72} and Rosen's \cite{Rosen71} two field variable 
theory with $\lambda=1$, yet both authors 
postulate a considerably more complex field equation than is used 
here.

\section{The Schwarzschild-like Solution}
\label{sec:SchwarzschildCase}

The PPN formulation predicts how the mid-range field of scalar 
gravity compares to general relativity, but it makes no such 
prediction about the strong-field region.  
A second measure of how ``good" the theory is, then, is to 
calculate the analytic results for a simple strong-field case and 
compare those results to the equivalent case in general relativity.  
The metric found in Section \ref{subsec:NewtonianLimit} 
corresponds to the well-studied Schwarzschild case of GR; in this 
section we will determine the properties of the equivalent case for 
scalar gravity, studying geodesics in a manner parallel to \S 25.3 of 
MTW \cite{MTW73}.

\subsection{The Effective Potential}
\label{subsec:EffectivePotential}
To begin, write (\ref{eq:SchwMetric}) in spherical coordinates
\begin{equation}
\label{eq:SphSchwMetric}
ds^{2}=-e^{-{2 M}/{r}}dt^{2}+e^{{2 M}/{r}} dr^{2} 
+ r^{2} e^{{2 M}/{r}} 
d\theta^{2} \eqperiod
\end{equation}
The geometry is unaffected by changes in $t$ or $\theta$, therefore 
two constants of motion exist:
\begin{equation}
\label{eq:MotionConstants}
p_{0} = -E \quad,\quad p_{\theta} = \pm L \eqperiod
\end{equation}
Conservation of rest-mass energy,
\begin{equation}
\label{eq:RestMassConsv}
g_{\alpha \beta}\, p^{\alpha} p^{\beta} + \mu^{2} = 0 \eqcomma
\end{equation}
gives the orbit equation
\begin{equation}
\label{eq:1stOrbitEq}
 -E^{2} e^{{2 M}/{r}} + e^{{2 M}/{r}} 
 \left(\!\frac{dr}{d\lambda}\!\right)^{2} 
 + \left(\frac{L}{r}\right)^{2} e^{-{2 M}/{r}} + 1 
= 0 
\end{equation}
where $\lambda$ is the path parameter and $p^{r} 
\equiv dr/d\lambda$.  
Defining 
$\tilde{E}={E}/{\mu}$, $\tilde{L}={L}/{\mu}$, 
$\lambda={\tau}/{\mu}$ places the orbit equation in its final 
form
\begin{equation}
\label{eq:OrbitEq}
\left(\frac{dr}{d\tau}\right)^{2} = \tilde{E}^{2} - V_{\rm eff}(r)^{2} \eqcomma
\end{equation}
where $V_{\rm eff}$, the effective potential, is defined as
%
%
\begin{figure}[t]
\label{fig:EffPot}
\begin{center}
\leavevmode
\epsfxsize=3.25in
\epsfbox{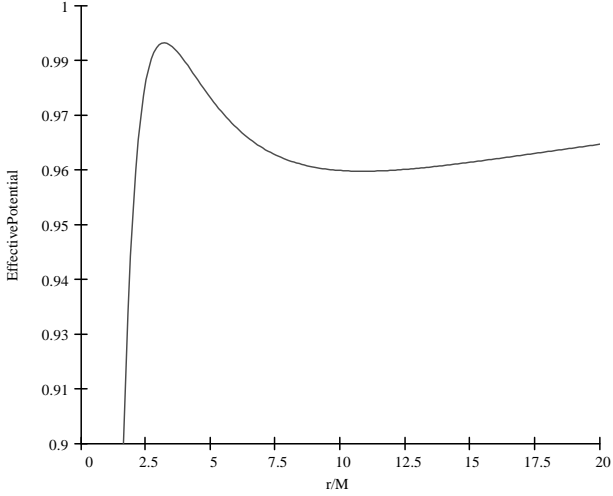}
\end{center}
\caption{The effective potential (\ref{eq:EffPot}) plotted as a 
function of $r/M$ for a typical specific angular momentum, in this
case $\tilde{L}=4 M$.
Notice the ``pit in the potential", (see MTW \protect \cite[Chap.25]
{MTW73})
which gives rise to an innermost stable circular orbit when, for
still smaller values of $\tilde{L}$, the stable and unstable
circular orbit radii (local min and max of $V_{\rm eff}$) coincide.
(see Section \ref{subsec:ISCO}).}
\end{figure}
%
%
\begin{equation}
\label{eq:EffPot}
 V_{\rm eff}(r)^{2} = e^{-{2 M}/{r}}\left(\frac{\tilde{L}^{2}}{r^{2}} 
 e^{-{2 M}/{r}} + 1\right) \eqperiod
\end{equation}
Figure 1 shows a typical plot of the effective potential 
as a function of $r$.

\subsection{Innermost Stable Circular Orbit}
\label{subsec:ISCO}

One of the first relativistic effects to be noticed was the ``pit" in the 
effective potential, the existence of which implies the presence of an 
innermost stable circular orbit.  
Any particle that moves within this limit must necessarily spiral 
in and impact the singularity at r=0.  
As the present theory also possesses this pit, it is reasonable 
to suppose that it possesses an innermost stable circular orbit as 
well.
 
Circular orbits occur at extrema of the effective potential.  
Taking the derivative with respect to $r$ of (\ref{eq:EffPot}) 
returns, after some algebra, the circular orbit angular momentum as 
a function of $r$:
\begin{equation}
\label{eq:CircAngMom}
\tilde{L}_{\rm circ} = r e^{{M}/{r}} \sqrt{\frac{M}{r-2 M}} \eqperiod
\end{equation}
It can immediately be seen that no circular orbits exist at all below 
$r = 2M$.  
The innermost stable circular orbit occurs at the value of $r$ for 
which $\tilde{L}_{\rm circ}$ is a minimum.  
Taking the derivative of (\ref{eq:CircAngMom}) with respect to $r$ 
yields
\begin{equation}
\label{eq:LastStableOrbit}
r_{\rm isco} = M (3 + \sqrt{5}) \eqperiod
\end{equation}
This result compares favorably to the predictions of general 
relativity.  
GR predicts the innermost stable circular orbit occurs at $r= 6 M$, 
which corresponds to a circumference of $12 \pi M$.  
The circumference of the innermost stable circular orbit under the 
scalar gravity theory is $12.6759 \pi M$, a difference of only 
$5.63\%$

Well inside the innermost stable circular orbit we cannot expect
this theory to mimic general relativity.    
To produce the analog of the interior Schwarzschild solutions 
would require inclusion of the $e^{-4\Phi} \delta_{ij}T^{ij}$ term 
in equation~\ref{eq:SCFFieldEq} and the hydrostatic fluid
equations, so it is possible that the theory would give a maximum
neutron star mass.
But since the radial null vectors are $k_{\pm} = \partial_t \pm 
e^{2\Phi} \partial_r$, the only possibility for a black hole 
effect would be to find solutions where $\Phi \rightarrow -\infty$ 
at a nonzero value of $r$.

\section{Gravitational Radiation}
\label{sec:GravRadiation}

With the completion of the LIGO gravitational wave interferometer 
nearing, attention in the numerical relativity community has been 
focused upon the production of gravitational wave templates to 
supply LIGO's matched-filtering analysis routines.  
A third and particularly relevant measure of how ``good" scalar 
gravity is would be to calculate the frequency of the radiation, as 
observed at infinity, of the innermost stable circular orbit.  

From the scalar gravity Schwarzschild-like metric
\begin{equation}
\label{eq:SphSchwMetric2}
ds^{2}=-e^{-{2 M}/{r}}dt^{2}+e^{{2 M}/{r}} dr^{2} 
+ r^{2} e^{{2 M}/{r}} (d\theta^{2} + 
\sin{\theta}^{2} d\phi^{2})
\end{equation}
the radiation frequency, $2\omega$, can be defined from the
circular orbit frequency 
\begin{equation}
\label{eq:OmegaDefn}
\omega = \frac{d\phi}{dt} = \frac{d\phi}{d\tau} \frac{d\tau}{dt} 
= e^{-{4 M}/{r}} 
\frac{\tilde{L}}{\tilde{E}r^{2}} \eqperiod
\end{equation}
Combining (\ref{eq:EffPot}) and (\ref{eq:CircAngMom}), yields the 
effective potential as a function of orbital radius; this value, by 
(\ref{eq:OrbitEq}), is also equal to the orbital energy squared.  
The result of this algebra is

\begin{equation}
\label{eq:OrbitalEnergy}
\tilde{E}_{\rm circ} =  e^{-{M}/{r_{\rm circ}}} 
\sqrt{\frac{r_{\rm circ}-M}{r_{\rm circ}-2M}} \eqperiod
\end{equation}

A somewhat more useful quantity conceptually, however, is the 
binding energy, 

\begin{equation}
\label{eq:BindingEnergy}
\tilde{E}_{\rm binding} = 1 - \tilde{E}_{\rm circ} \eqperiod
\end{equation}
Substituting (\ref{eq:OrbitalEnergy}) and (\ref{eq:CircAngMom}) 
into (\ref{eq:OmegaDefn}) gives, after some algebra, scalar 
gravity's equivalent of Kepler's Third Law
\begin{equation}
\label{eq:Kepler3rd}
\omega^{2} r_{\rm circ}^{2}e^{{4 M}/{r_{\rm circ}}} (r-M) = M 
\eqperiod
\end{equation}
%
%
\begin{figure}[t]
\label{fig:FreqSpectrum}
\begin{center}
\leavevmode
\epsfxsize=3.25in
\epsfbox{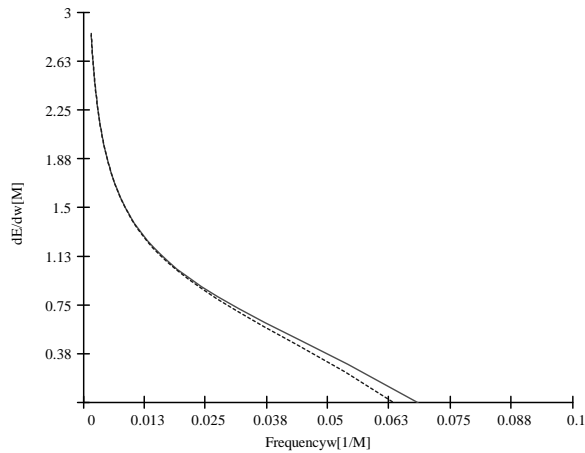}
\end{center}
\caption{The frequency spectrum of gravitational radiation 
produced by a test mass in nearly circular orbit about a 
central object as the test mass adiabatically spirals inward.
The graph shows the change in binding energy (\ref{eq:BindingEnergy})
as a function of circular orbit frequency (\ref{eq:OmegaDefn}), which 
is in turn a function of the circular orbit radius.  
The solid line shows the spectrum for a test mass moving within a 
Schwarzschild background metric, while the dashed line shows the 
spectrum for a test mass moving within the background metric of 
(\ref{eq:SphSchwMetric2}).  
The innermost stable circular orbit occurs where each curve crosses 
the $x$-axis.  
The agreement between the curves is very good almost down to this 
point.}
\end{figure}
%
%
Notice that this expression is precisely Kepler's Third Law only for 
large ${r}/{M}$. 
Although $\tilde{E}_{\rm binding}(\omega)$ cannot be found as an
analytic expression, we can easily plot the spectrum of the
gravitational radiation $d\tilde{E}_{\rm binding}(\omega)/d\omega$ 
by knowing both $d\tilde{E}_{\rm binding}(\omega)/d\omega =
(d\tilde{E}_{\rm binding}(\omega)/dr_{\rm circ})/
(d\omega/dr_{\rm circ})$ and $\omega$ as functions of 
$r_{\rm circ}$. 
Figure 2 compares the frequency spectrum 
calculated under scalar gravity with that found from general 
relativity.  
As can be seen, the two theories agree quite well almost down to 
the innermost stable circular orbit.  
The frequency of the innermost stable circular orbit is found using 
(\ref{eq:Kepler3rd}) evaluated at $r = r_{\rm isco} = M (3 + \sqrt{5})$:
\begin{equation}
\label{eq:ISCOomega}
\omega_{\rm isco} = \frac{0.0633326}{M} \eqperiod
\end{equation}
General relativity gives a value of $\omega_{\rm isco} 
= 0.0680414 / M$ (a difference of $6.92\%$), again 
demonstrating the close agreement the two theories.
This close agreement is not too surprising when one considers 
that there is a sense in which the Schwarzschild-like solution 
of this SCF scalar theory ``almost'' satisfies Einstein's equations.
The non-zero components of the Einstein tensor, calculated using 
the metric (\ref{eq:SphSchwMetric2}) are
\begin{equation}
\label{eq:SGEinstein}
 G^{0}_0 = -G^{1}_1 = G^{2}_2 = G^{3}_3 =
 e^{-2M/r}\frac{M^2}{r^4}
\eqcomma
\end{equation}
rather than zero.
This is smaller than the nonzero
components of the Riemann tensor in the Schwarzschild metric 
by a factor $e^{-2M/r}(M/r)$.
A dimensionless measure of this failure to satisfy
Einstein's equations has its maximum at $r=M$ where each
nonzero component $M^2 |G^\mu_\nu|$ is $e^{-2} = 0.135$, from 
which point they decrease monotonically.  At $r = r_{\rm isco}$ 
this error is $0.0009$ and decreases rapidly farther out.

\section{Discussion}
\label{sec:Discussion}

The scalar theory presented here is intended as a stand-in for
general relativity while developing hydrodynnamic computer codes
for application in problems such as the production of
gravitational radiation by binary neutron star systems.
It has the advantage of involving only one field variable $\Phi$
(and its time derivative) and satisfying a simple field equation.
The field equation (\ref{eq:SCFFieldEq}) involves only the usual
flat spacetime d'Alembertian (nonlinearities only in terms
containing no derivatives) and reduces to the flat space 
linear wave equation outside the region occupied by matter.
This should simplify its numerical treatment, and particularly 
the imposition of outgoing wave boundary conditions; there 
are no curved light cones to manage.  
Its interface to hydrodynamics can be identical that of general 
relativity --- it requires a matter stress-energy tensor and 
produces a metric.

It seems plausible that, in a range including fields stronger than
those treated by the PPN methods, this theory could also give
results that are close to those of general relativity.
For Coulomb-like gravitational fields our study of the analog of
the Schwarzschild metric shows results within a few percent of
those of GR down to the innermost stable circular orbit.
For gravitomagnetic forces (from metric components $g_{0i}$
generated by matter currents $\rho v$) it will necessarily fail,
as it allows no $g_{0i}$ metric components. 
Thus one cannot expect physically plausible results when matter 
velocities approach the velocity of light.  
But since this scalar theory does reproduce important
qualitative features of general relativity (the existence of an
innermost stable circular orbit for test particles), it may 
allow the initial study of important relativistic hydrodynamical 
phenomena whose accurate analysis will have to be done using the
full Einstein theory of gravitation.

\acknowledgments
We thank Conrad Schiff for many helpful conversations. 
We thank Joan Centrella and Richard Matzner
for valuable discussions in the early stages of formulating
this research program. 
This research was supported in part by 
NSF grant PHY 9700672.


\begin{references}

\bibitem[*]{PP}UMD PP-00-029; gr-qc/9910032

\bibitem{Fox96}G. Fox, Syracuse University CPS713 Module on 
Numerical Relativity, p. 11 (1996). 

\bibitem{SeidSu99}Edward Seidel and Wai-Mo Suen, gr-qc/9904014.

\bibitem{ST93}S. L. Shapiro and S. A. Teukolsky, Phys. Rev. D 
{\bf 47}, 1529 (1993). 

\bibitem{MTW73}C. W. Misner, K. S. Thorne, and J. A. Wheeler, 
{\it Gravitation}, (Freeman, San Francisco, 1973). 

\bibitem{Ni72}W. Ni, Ap.J. {\bf 176}, 769 (1972).

\bibitem{ThorneWill71}K. S. Thorne and C. M. Will, Ap. J. 
{\bf 163}, 595 (1971).

\bibitem{Ni73}W. Ni, Ap. J. {\bf 181}, 939 (1973).

\bibitem{NewWatt98}K.New, K.Watt, C.W.Misner, J.Centrella, 
Phys. Rev. D, {\bf 58}, 064022 (1998).

\bibitem{Hill51}E.L.Hill, Rev. Mod. Phys. {\bf 23}, 253 (1951).

\bibitem{Byers98}N.Byers, in {\it The Heritage of Emmy Noether, 
Israel Mathematical Conference Proceedings, Vol. 12}, Min
Teicher editor, (Bar Ilan University, Tel Aviv, Israel, 1999,
distributed by Am. Math. Soc.); \\
http://xxx.lanl.gov/ps/physics/9807044 .

\bibitem{LL71}Landau and Lifshitz, {\it The Classical Theory of 
  Fields}, Third English edition, (Pergamon, 1971), equation 32.5.

\bibitem{Ryder85}Lewis H. Ryder, {\it Quantum Field Theory}, 
(Cambridge University Press, Cambridge, 1985) 
Section 3.2, equation 3.20.

\bibitem{PeskSch95}Michael E. Peskin and Daniel V. Schroeder, 
{\it An Introduction to Quantum Field Theory}, 
(Addison-Wesley Publishing Company, New York, 1995)
Chapter 2, section 2.2 equation 2.17.

\bibitem{JimenezVazquez90}S. Jim\'{e}nez and L. V\'{a}zquez,
Appl. Math \& Comp. {\bf 35}, 61 (1990).

\bibitem{NordtvedtWill72}K. Nordtvedt, Jr., and C.M.Will, Ap. J. 
{\bf 177}, 775 (1972).

\bibitem{Rosen71}N. Rosen,  Phys. Rev. D, {\bf 3}, 2317 (1971).

\end{references}
\end{document}